\documentclass[aps,twocolumn,prl]{revtex4}
\usepackage{graphicx}
\usepackage{amsfonts}
\usepackage{amssymb}
\usepackage{amsmath}

\begin{document}

\title{\bf Dark matter-wave solitons in the dimensionality crossover\\
}

\author{
G.\ Theocharis$^{1}$, 
P.G.\ Kevrekidis$^{2}$, 
M.K.\ Oberthaler$^{3}$ and 
D.J.\ Frantzeskakis$^{1}$ 
}
\affiliation{
$^{1}$ Department of Physics, University of Athens, Panepistimiopolis,Zografos, Athens 157 84, Greece \\
$^{2}$ Department of Mathematics and Statistics,University of Massachusetts, Amherst MA 01003-4515, USA\\ 
$^{3}$ Kirchhoff Institut f\"{u}r Physik, INF 227, Universit\"{a}t Heidelberg, 69120, Heidelberg, Germany}

\begin{abstract}

We consider the statics and dynamics of dark matter-wave solitons in the dimensionality crossover regime from 3D to 1D. 
There, using the nonpolynomial Schr\"{o}dinger mean-field model, we find that the anomalous mode of the Bogoliubov spectrum has an 
eigenfrequency which coincides with the soliton oscillation frequency obtained by the 3D Gross-Pitaevskii model. 
We show that substantial deviations (of order of $10\%$ or more) from the 
characteristic frequency $\omega_{z}/\sqrt{2}$ ($\omega_{z}$ being the 
longitudinal trap frequency) are possible even in the purely 1D regime. 
\end{abstract}

\maketitle 

The experimental realization of lower-dimensional Bose-Einstein condensates (BECs) 
in highly anisotropic traps \cite{bec1da,bec1db} has inspired many studies devoted to the behavior of such systems in the 
dimensionality crossover, i.e., from three-dimensions (3D) to one-dimension (1D). Particularly, it has 
been shown that fundamental 
properties of BECs, such as the chemical potential, 
speed of sound, 
collective oscillations, change significantly as the dimensionality is reduced 
from 3D to 1D \cite{beh,str}. These regimes, as well as the {\it crossover} between them, can be described \cite{str} by 
the dimensionless parameter $N\Omega\alpha/\alpha_{\perp}$, where $N$ is the number of atoms, 
$\Omega=\omega_z/\omega_{\perp}$ is the ratio of longitudinal and transverse trapping frequencies, $\alpha$ is the 
scattering length, and $\alpha_{\perp}=\sqrt{\hbar/m \omega_{\perp}}$ ($m$ is the atomic mass). 
In particular, if the dimensionality parameter is $N\Omega\alpha/\alpha_{\perp}\gg1$, the BEC locally 
retains its original 3D character 
and its ground state can be described by the Thomas-Fermi approximation in all directions. On the other hand, if $N\Omega\alpha/\alpha_{\perp} \ll 1$, 
excited states along the transverse direction are not energetically 
accessible and the BEC is effectively 1D; in 
such a case, the transverse wave function is described by the 
ground state of the radial harmonic oscillator, 
whereas the longitudinal wave function is governed by an effectively 1D Gross-Pitaevskii equation (GPE) \cite{gpe1d}. 
Of particular relevance are effectively 1D mean-field models \cite{kam,npse} that have been developed to describe 
the axial dynamics of 
``cigar-shaped'' BECs in the 1D (and 3D) regimes, as well as 
in the dimensionality crossover. Importantly, there exist recent experimental results \cite{markus1} 
which have been accurately described by the model of \cite{npse}.

It is then natural to expect that, apart from the ground-state properties of 
BECs, the dimensionality should significantly affect the 
stability and dynamical properties of the {\it excited} BEC states as well, such as the dark solitons in repulsive BECs. 
The first experiments reporting the observation of dark matter-wave solitons were performed with quasi-spherical \cite{nist,bpa} 
or cigar-shaped \cite{han} traps. In all cases, dark solitons were prone to instabilities, such as non-uniformity 
induced dynamical instability (leading to a U-shaped deformation of the soliton propagation front) 
\cite{nist}, 
thermal instability (decay due to the interaction with the thermal cloud) 
\cite{han}, or snaking instability (leading to soliton decay into vortex rings) \cite{bpa}. 
From the theoretical point of view, the former two instabilities were 
analyzed in \cite{Mur2}, while the snaking instability (which occurs in the higher-dimensionality setting \cite{kuz}) 
was analyzed in \cite{Muryshev,mprizolas,brand}. In the latter works, a detailed study of the Bogoliubov-de Gennes (BdG) 
equations in 2D and 3D revealed the emergence of complex eigenvalues in the 
excitation spectrum and their connection 
to oscillatory dynamical instabilities (including the snaking instability). Moreover, it was demonstrated \cite{Muryshev,mprizolas,brand,dz} 
that there exists a so-called ``anomalous mode'' in the excitation spectrum, which has negative energy.
The existence of the anomalous mode indicates that dark solitons 
(and vortices \cite{Fetter}) are thermodynamically unstable and, in the 
presence of dissipation, the system is driven towards configurations with 
lower energy; this scenario is also often referred to as energetic 
instability \cite{wuniu}.
Importantly, the anomalous mode frequency is directly related to the dark soliton oscillation frequency 
$\omega_{\rm osc} = \omega_{z}/\sqrt{2}$ \cite{Muryshev,dz}. This oscillation frequency has been obtained 
upon analyzing the 1D GPE using different analytical approaches \cite{oscfreq} for $\Omega \ll 1$, 
and assuming the Thomas-Fermi (TF) approximation 
for the longitudinal BEC background on which the dark soliton exists. 
However, it is important to stress that the above result is {\it not} 
particularly relevant 
to either the first dark soliton experiments \cite{nist,bpa,han}, or to the recent experimental studies of \cite{last_exp}, 
because of the fact that, in all cases, the condensates were actually in the purely 3D regime.

In this work, we revisit this problem and study 
the statics and dynamics of dark solitons in BECs confined in highly elongated 
traps (with $\Omega\ll 1$) in the dimensionality crossover 
regime, i.e., for $N\Omega\alpha/\alpha_{\perp}\approx 1$; our analysis also applies in the 
case of quasi-1D {\it small} BECs in which the TF approximation for the axial direction is {\it not} valid.  
We thus aim to provide a complete picture bridging the one- and multi-dimensional
dynamics of dark solitons. 
We will show, in particular, that in the crossover regime the anomalous mode frequency 
resulting from the BdG analysis of the nonpolynomial Schr\"{o}dinger equation (NPSE) \cite{npse} 
estimates accurately the soliton oscillation frequency resulting from the 3D GPE. Moreover, we will demonstrate that, in this regime, 
the frequencies of the anomalous mode and the soliton oscillation coincide but may significantly differ from the value $\omega_{z}/\sqrt{2}$ 
(deviations may easily exceed $10\%$). 

Let us start our analysis considering the example of a $^{87}$Rb condensate containing $N \sim 2100$ atoms and  
confined in a trap with frequencies $\omega_{\perp}= 20 \omega_{z} = 2\pi \times 200$ Hz (here $\Omega=0.05$). In this case, 
the dimensionless parameter $N\Omega \alpha/\alpha_{\perp} = 0.815$, a value 
in the dimensionality crossover region, and thus 
the 1D GPE is not valid (the transverse wave function deviates from the ground state of the transverse harmonic oscillator). 
Thus, to calculate the oscillation frequency of a dark soliton in such a setting, one needs to 
consider the fully 3D GPE, which can be expressed in the following dimensionless form:
\begin{equation}  
i\frac{\partial \psi}{\partial t}=\left[-\frac{1}{2}\nabla^2 + \frac{1}{2}(\Omega^2 z^2 + r^2) + 
|\psi|^2\right]\psi, 
\label{3dGPE}
\end{equation}
where the density $|\psi|^2$, length, time and energy are respectively measured in units of 
$4\pi a a_{\perp}^{2}$, $\alpha_{\perp}$, $\omega_{\perp}^{-1}$ and $\hbar\omega_{\perp}$. 
We use as an initial condition a dark soliton initially placed at $z=2$ on top of the ground state 
of the system, which is found by a relaxation technique (i.e., using 
imaginary time integration to find 
the ground state, as well as a quiescent dark soliton at the trap center, and then displacing the soliton 
at the desired location to set the soliton into motion).
Then, for a normalized 3D chemical potential 
$\mu_{3D}=1.625$, the following result is obtained (see Fig. 1): The soliton oscillates with a frequency 
$\omega_{\rm osc}=0.746 \Omega$, although the system is ``highly anisotropic'' and one would expect 
an oscillation frequency of $\Omega/\sqrt{2}=0.707 \Omega$ (in this case the deviation from the prediction 
of \cite{oscfreq} is of order of $\approx 6\%$).

\begin{figure}[tbp]
\includegraphics[width=8cm]{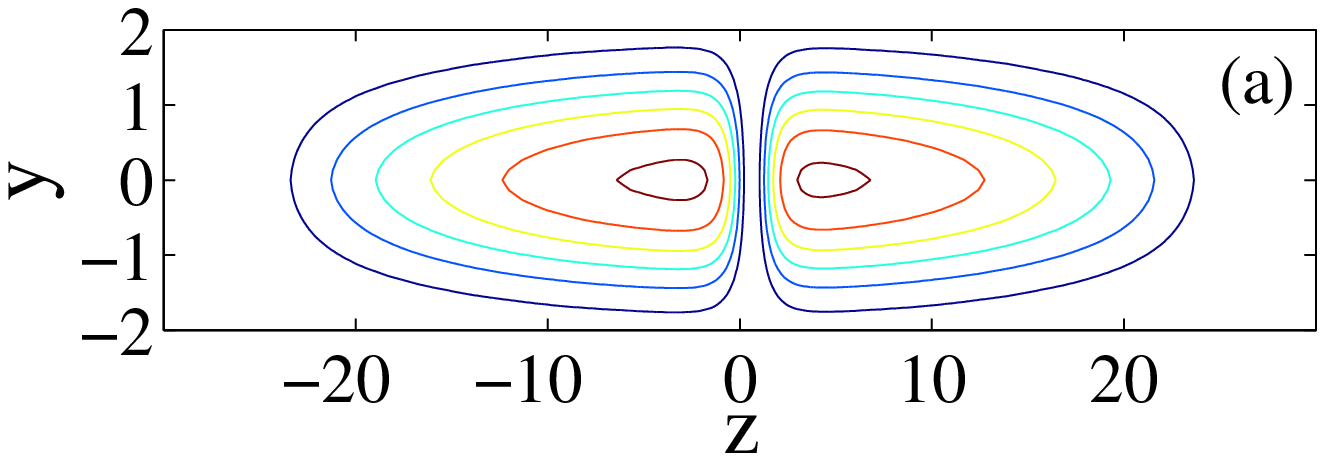}
\includegraphics[width=8cm]{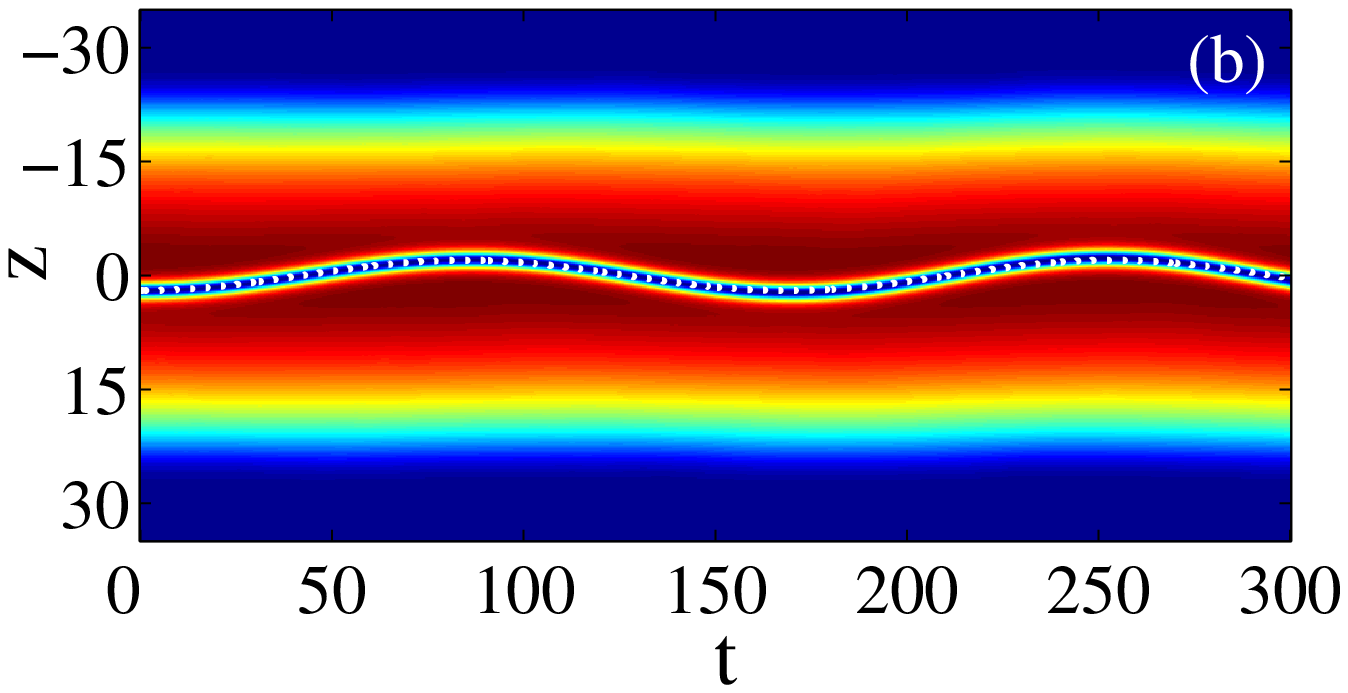}
\caption{(Color online) (a) Contour plot of the initial density (in the $x=0$ plane) of a BEC, 
confined in an harmonic trap with $\Omega=0.05$, and with a dark soliton placed at $z=2$. 
The dimensionality parameter is $N\Omega\alpha/\alpha_{\perp}=0.815$. 
(b) Spatio-temporal evolution of the above density along a cross-section at $r=0$. 
The dark soliton oscillates with frequency $\omega_{\rm osc}=0.746\Omega>\Omega/\sqrt{2}$. 
The dotted line across the soliton trajectory corresponds to the prediction obtained by the NPSE model.
}
\label{fig1}
\end{figure}

Instead of using the fully 3D GPE model, it is worth investigating if the above value of the soliton oscillation frequency 
can be obtained in the framework of a 1D mean field model. As the 1D GPE 
is {\it not} applicable in this example, 
we consider another relevant candidate, namely the NPSE of Ref. \cite{npse}. The latter, can be expressed in the 
following dimensionless form, 
\begin{equation}  
i\frac{\partial \phi}{\partial t}=\left[-\frac{1}{2}\frac{\partial^2}{\partial z^2} + \frac{1}{2}\Omega^2 z^2 + 
\frac{3|\phi|^2+2}{2(1+|\phi|^2)^{1/2}}\right]\phi, 
\label{1dNPSE}
\end{equation}
where $\phi$ is the normalized longitudinal wavefunction, and the density $|\phi|^2$ is measured in units of $2aN$ 
(length, time and energy are measured in the same units as in the 3D GPE). Note that this model is reduced to the 
1D GPE in the low density limit of $|\phi|^2 \ll 1$ (in our units), which is not relevant to the case example of 
$N\Omega\alpha/\alpha_{\perp} =0.815$. We have numerically 
integrated Eq. (\ref{1dNPSE}) with an initial condition corresponding to longitudinal density profile of the 3D GPE case and have found that 
the soliton oscillation frequency is $\omega_{\rm osc}=0.744 \Omega$, approximately the same with the one obtained by the 
3D GPE ($\omega_{\rm osc}=0.746\Omega$). This result, 
depicted by the dotted line in Fig. \ref{fig1}(b), 
indicates that the NPSE may accurately predict the soliton oscillation frequency in the dimensionality crossover regime.

\begin{figure}[tbp]
\includegraphics[width=3.85cm,height=3.45cm]{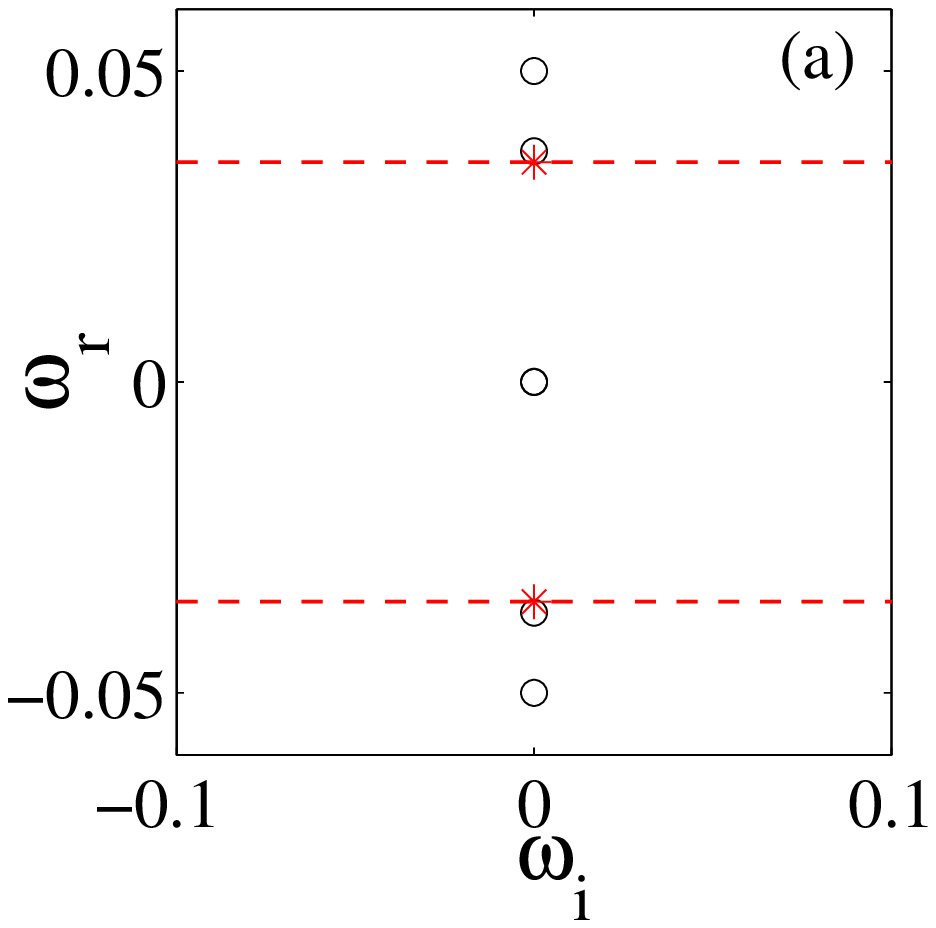}
\includegraphics[width=4.5cm,,height=3.4cm]{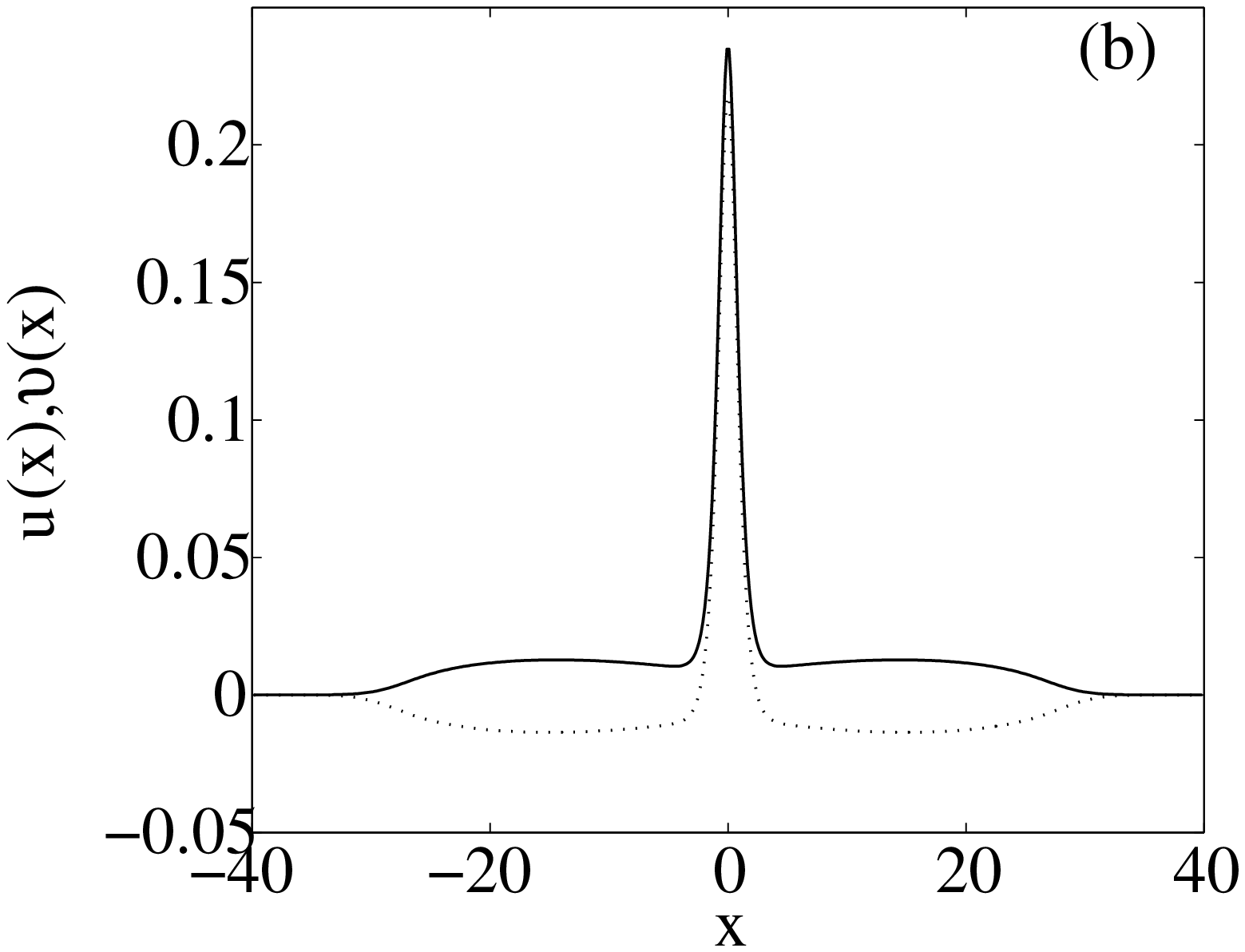}
\caption{(Color online)(a) The six lowest characteristic eigenfrequencies of the Bogoliubov excitation
spectrum: Two of them, located at the origin, correspond to the 
Goldstone mode due to the phase invariance, 
the ones at $\pm \Omega = \pm 0.05$ to the Kohn mode, 
while the indermediate ones, at $\pm 0.744\Omega$, to the anomalous mode. The 
dashed lines indicate the location of the $\pm \Omega/\sqrt{2}$ prediction.
(b) The eigenfunctions $u$ (solid line) and $\upsilon$ (dotted line) of the anomalous mode.}
\label{fig3}
\end{figure}

Following the above argument, we now proceed with the investigation of the excitation spectrum of the NPSE's dark soliton state.
This way, we first find a real, stationary soliton state $\phi_{\rm ds}(z)$ using a fixed point algorithm (a standard Newton-Raphson method \cite{atkinson}). 
Then, considering small perturbations of this state of the form,  
\begin{equation}
\phi(z,t)=\left[\phi_{\rm ds}(z)+\epsilon\left(u(z)e^{-i\omega t}+\upsilon^{\ast}(z)e^{i\omega t}\right)\right]e^{-i\mu t},
\label{ansatz}
\end{equation}
we obtain from Eq. (\ref{1dNPSE}) the following BdG equations (valid to 
leading order in the small parameter $\epsilon$): 
\begin{eqnarray}
\omega u &=& [\hat{H} - \mu + f(\phi_{ds}^{2})] u + g(\phi_{ds}^{2})\upsilon, 
\\
-\omega \upsilon &=& [\hat{H} - \mu + f(\phi_{ds}^{2})] \upsilon + g(\phi_{ds}^{2})u, 
\label{BdG}
\end{eqnarray}
where $\hat{H}= -(1/2)\partial_{z}^{2}+(1/2)\Omega^2 z^2$, 
$f(\phi_{\rm ds}^2)=\frac{9\phi_{\rm ds}^4+14\phi_{\rm ds}^2+4}{4 \left(1+\phi_{\rm ds}^2\right)^{3/2}}$ and 
$g(\phi_{\rm ds}^2)=\frac{3\phi_{\rm ds}^4 + 4\phi_{\rm ds}^2}{4\left(1+\phi_{\rm ds}^2\right)^{3/2}}$.
The above equations provide the eigenfrequencies $\omega \equiv \omega_{r}+i \omega_{i}$ and the amplitudes $u$ and $\upsilon$ of the normal modes
of the system. Note that due to the Hamiltonian nature of the system, 
the eigenfrequencies of the Bogoliubov analysis appear in pairs (or in quartets, if complex); thus, 
the solution of BdG equations with frequency $\omega$ represent the same physical oscillation with the solution with 
frequency $-\omega$ \cite{book}. 

Among the various eigenfrequencies, we focus on the three smallest magnitude pairs
in Fig. \ref{fig3}(a): 
One of them is at the origin of the spectral plane $(\omega_{r}, \omega_{i})$ 
reflecting the phase invariance of the NPSE model.  
The respective eigenfunction 
is the so-called Goldstone mode 
and does not result in any physical excitation (oscillation) of the system. 
On the other hand, the solutions with eigenfrequencies 
$\omega= \Omega= 0.05$ correspond to the so-called dipole (or Kohn) mode, 
representing the motion of the center of mass of the system 
which oscillates with the frequency of the harmonic trap. 
Finally, of particular interest are the solutions with eigenfrequencies $\omega= 0.744 \Omega$ corresponding to 
the so-called {\it anomalous mode}. For the latter, the integral of the norm$\times$energy product, 
$\int \omega (|u|^2 -|v|^2) dz $ (in our units), is negative \cite{book}, indicating the energetic instability 
of the dark soliton discussed above. 
Importantly, this eigenfrequency coincides with the oscillation frequency of the dark soliton 
as obtained by the NPSE model--and in accordance with the 
result of the 3D GPE. The eigenfunctions $u$ and $\upsilon$ of the anomalous mode are shown in Fig. \ref{fig3}(b). As seen, they are 
localized within the notch of the dark soliton \cite{dz}.

\begin{figure}[tbp]
\includegraphics[width=8cm]{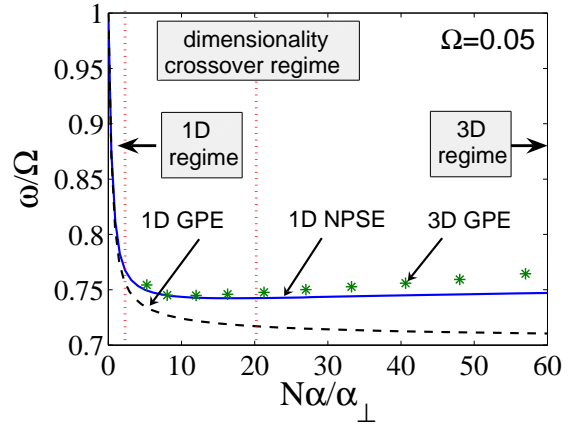}
\caption{(Color online) Normalized frequency $\omega/\Omega$ as a function of $N\alpha/\alpha_{\perp}$: 
Solid (dashed) line corresponds to the anomalous mode frequency obtained by the BdG analysis of the NPSE (1D GPE), and 
stars denote the oscillation frequency of the dark soliton obtained by the 3D GPE. The vertical dotted lines 
depict the different dimensionality regimes.
}
\label{fig4}
\end{figure}

We now aim to investigate if the above agreement is generic, i.e., whether the BdG analysis of the NPSE model indeed provides an 
accurate estimation of the soliton oscillation frequency in the dimensionality crossover regime. In this 
connection, we fix the normalized harmonic trap strength to its previous value ($\Omega = 0.05$), and find the soliton 
oscillation frequency varying the parameter $N \alpha/\alpha_{\perp}$. 
The results are presented in Fig. \ref{fig4} which summarizes the main
findings of the present work. The obtained normalized 
frequency $\omega/\Omega$ is shown: the solid line depicts the anomalous mode frequency obtained 
by the BdG analysis of the NPSE model, while the stars depict the soliton oscillation frequency obtained 
by a direct numerical integration of the fully 3D GPE. The different dimensionality regimes, 
are also shown in Fig. \ref{fig4}, and are defined as follows: The 1D regime corresponds to 
$\mu_{3D} < \hbar \omega_{\perp}$ \cite{bec1da}, or $\mu_{3D} < 1$ in our units, i.e., to 
$N \alpha/\alpha_{\perp} < 2$ (or $N \Omega \alpha/\alpha_{\perp} < 0.1$) 
as per our normalizations; on the other hand, the dimensionality crossover 
regime corresponds to the region around $N \alpha/\alpha_{\perp} \approx 20$ (or $N \Omega \alpha/\alpha_{\perp} \approx 1$), 
while the 3D regime corresponds to the limit $N \alpha/\alpha_{\perp} \gg 20$ (or $N \Omega \alpha/\alpha_{\perp} \gg 1$).

In the 1D regime, 
the NPSE model is reduced to the 1D GPE and, as a 
result, the anomalous mode frequency obtained in the framework of the 
NPSE coincides with the one obtained by the BdG analysis of the 
1D GPE (see dashed line in Fig. \ref{fig4}). 
This agreement ceases to exist for $N \alpha/\alpha_{\perp} > 2$, which is a clear indication 
that the system enters the dimensionality crossover regime. 
Therefore, the asymptotic value of $\Omega/\sqrt{2}$ depicted by the dotted line in Fig. \ref{fig4} 
for $N \alpha/\alpha_{\perp} \gg 2$ is {\it quantitatively irrelevant}: The correct result 
for the oscillation frequency in the crossover regime is 
provided by the 3D GPE (stars) and is accurately
approximated by the NPSE (solid line).
Notice that in the limit $N \alpha/\alpha_{\perp} \rightarrow 0$ (which corresponds to the linear 
Schr\"{o}dinger equation), one obtains $\omega/\Omega \rightarrow 1$, i.e., the anomalous and Kohn mode frequencies 
coincide, in accordance to the prediction of Ref. \cite{mprizolas}. 

In the dimensionality crossover regime, and particularly in the interval 
$2 \le N \alpha/\alpha_{\perp} \le 20$, the anomalous mode frequency obtained by 
the BdG analysis of the NPSE almost coincides with the oscillation frequency of the dark soliton obtained by the 
3D GPE. However, as shown in Fig. \ref{fig4}, as the system approaches the 3D regime ($N \alpha/\alpha_{\perp} \gg 20$) 
the NPSE underestimates the frequency obtained by the 3D GPE. This is a consequence of the fact that the NPSE 
uses a gaussian ansatz to describe the transverse wavefunction, rather than the TF profile which is relevant to the 3D regime. On the 
other hand, it is important to note that for $\mu_{3D}>2.4$ \cite{Muryshev,mprizolas}, corresponding to $N \alpha/\alpha_{\perp} \approx 60$, 
the system becomes dynamically unstable through the emergence of complex eigenfrequencies in the excitation spectrum of the 
3D GPE, related to the onset of oscillatory and snaking instabilities of dark solitons \cite{Muryshev,mprizolas,brand}. 
Although these types of instabilities were indeed observed in our 3D simulations (results not shown here), 
the BdG equations of the NPSE did not predict such complex eigenvalues. Thus, the above results suggest a note of caution: the BdG analysis of the 
NPSE model can indeed predict accurately the oscillation frequency of 
dark solitons in the dimensionality crossover regime, but, given its
freezing of the transverse directions into their ground state, 
it {\it cannot} capture the emergence of the pertinent 
instabilities occuring in the {\it fully} 3D regime. 

\begin{figure}
\includegraphics[width=9cm]{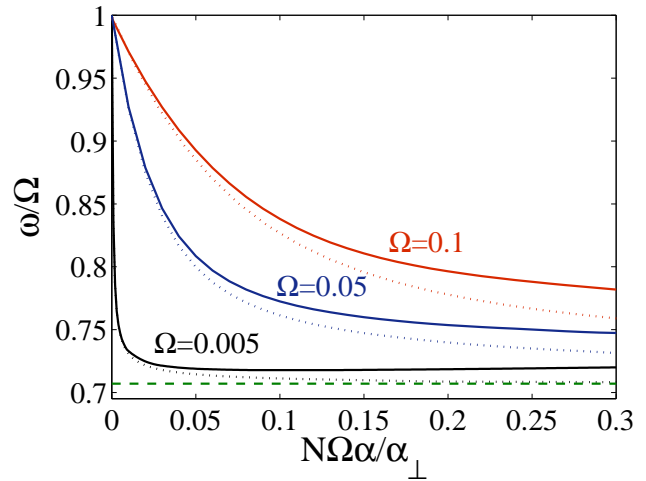}
\caption{(Color online) Normalized frequency $\omega/\Omega$ of the anomalous mode obtained 
by BdG analysis of the NPSE (solid lines) and the 1D GPE (dotted lines), as a function of 
$N\Omega\alpha/\alpha_{\perp}$ for $\Omega=0.1, 0.05, 0.005$. The dashed line indicates the $\Omega/\sqrt{2}$ anomalous mode eigenfrequency.}
\label{fig5}
\end{figure}

We have also performed the BdG analysis of both the NPSE and 1D GPE models for other values of the normalized trap frequency $\Omega$. 
As seen in Fig. \ref{fig5}, where $\omega/\Omega$ is shown as a function of the dimensionality parameter $N \Omega \alpha/\alpha_{\perp}$, 
larger (smaller) values of $\Omega$ yield larger (smaller) anomalous mode frequencies. It is worth mentioning that Fig. \ref{fig5} 
is merely devoted to the 1D regime (the dimensionality parameter does not exceed the value $0.3$). 
There, as the trap frequency is decreased (towards the limit of $\Omega \ll 1$ considered in Refs. \cite{oscfreq}), 
the anomalous mode frequency approaches the value of $\Omega/\sqrt{2}$ depicted by the dashed line in Fig. \ref{fig5}. Thus, in the 1D 
regime, and for $\Omega \ll 1$, the BdG analysis of the NPSE recovers the 
soliton oscillation frequency obtained asymptotically
in the framework of the 1D GPE \cite{oscfreq}. 

An important observation stemming from our analysis is that even in the 1D regime, the soliton oscillation frequency may have a 
substantial difference from $\Omega/\sqrt{2}$. For example, let us consider two different cases (both for a $^{87}$Rb BEC), one with  
number of atoms $N \sim 5000$ and trap frequencies $\omega_{\perp}=200\omega_{z}= 2\pi \times 200$Hz, and one with  
$N\sim 1000$ and $\omega_{\perp}=10\omega_{z}= 2\pi \times 70$Hz. In the former case, our analysis predicts that the soliton oscillation 
frequency will be $\omega_{\rm osc} = 0.718 \Omega$, differing only $1.5\%$ from the value of $\Omega/\sqrt{2}$, while in the second case 
$\omega_{\rm osc} = 0.772 \Omega$, differing $10\%$ 
from 
$\Omega/\sqrt{2}$. 
Such strong deviations from the asymptotic prediction should be directly accessible to current experimental settings.

In conclusion, we have studied dark-matter wave solitons in Bose-Einstein
condensates and how their dynamics is affected by dimensionality in 
settings of experimental interest/accessibility.
We have first shown that in a highly anisotropic system at the crossover
between 3D and 1D behavior, the dark-soliton oscillation frequency is larger than the 
usually quoted value of $\Omega/\sqrt{2}$. The deviations are predicted 
by the 1D mean-field NPSE model, whose excitation spectrum reveals an 
anomalous mode eigenfrequency identical to the soliton oscillation frequency in the dimensionality crossover regime. Limitations of the 
NPSE mdel, concerning its validity towards the fully 3D regime (where dynamical instabilities of dark solitons may also come into play) were also
discussed. Importantly, our analysis demonstrates that, even in the purely 1D regime, deviations of the soliton oscillation frequency 
from the standardly used theoretical  value 
of $\Omega/\sqrt{2}$ of order of $10\%$ (or even more) 
are possible and should be observable in current experimental setups. 

This work has been supported from ``A.S. Onasis'' 
Foundation (G.T.), the Special Research Account of the University of Athens (G.T., D.J.F.), 
and NSF-DMS-0204585, NSF-DMS-0505663, NSF-0619492 and NSF-CAREER (P.G.K.).



\begin{thebibliography}{99}

\bibitem{bec1da} A. G\"{o}rlitz {\it et al.}, Phys. Rev. Lett. {\bf 87}, 130402 (2001).

\bibitem{bec1db} W. H\"{a}nsel {\it et al.}, Nature (London) {\bf 413}, 498 (2001); H. Ott {\it et al.}, Phys. Rev. Lett. {\bf 87}, 230401 (2001).

\bibitem{beh} K. K. Das, Phys. Rev. A {\bf 66}, 053612 (2002).

\bibitem{str} C. Menotti and  S. Stringari, Phys.\ Rev.\ A \textbf{66}, 043610 (2002).

\bibitem{gpe1d} V. V. P\'{e}rez-Garc\'{i}a, H. Michinel, and H. Herrero, Phys. Rev. A {\bf 57}, 3837 (1998).

\bibitem{kam} A. M. Kamchatnov and V. Shchesnovich, Phys. Rev. A {\bf 70}, 023604 (2004).

\bibitem{npse} L. Salasnich, A. Parola, and L. Reatto, Phys.\ Rev.\ A \textbf{65}, 043614 (2002).

\bibitem{markus1} M. Albiez {\it et al.}, Phys. Rev. Lett. {\bf 95}, 010402 (2005).

\bibitem{nist} J. Denschlag {\it et al.}, Science {\bf 287}, 97 (2000).

\bibitem{bpa} B. P. Anderson {\it et al.}, Phys. Rev. Lett. {\bf 86}, 2926 (2001).

\bibitem{han} S. Burger {\it et al.}, Phys. Rev. Lett. {\bf 83}, 5198 (1999).

\bibitem{Mur2} A. Muryshev \textit{et al.}, Phys.\ Rev.\ Lett. \textbf{89}, 110401 (2002).

\bibitem{kuz} E. A. Kuznetsov and S. K. Turitsyn, Zh. Eksp. Teor. Fiz. {\bf 94}, 119 (1988) 
[Sov. Phys. JETP {\bf 67}, 1583 (1988)].

\bibitem{Muryshev} A. E. Muryshev, H. B. van Linden van den Heuvel, and G. V. Shlyapnikov, 
Phys.\ Rev.\ A \textbf{60}, R2665 (1999).

\bibitem{mprizolas} D. L. Feder \textit{et al.}, Phys.\ Rev.\ A \textbf{62}, 053606 (2000).

\bibitem{brand} J. Brand and W. Reinhardt, Phys.\ Rev.\ A \textbf{65}, 043612 (2002).

\bibitem{dz} J. Dziarmaga and K. Sacha, Phys. Rev. A {\bf 66}, 043620 (2002).

\bibitem{Fetter} A. L. Fetter and A. A. Svidzinsky, J. Phys.: Condens. Matter \textbf{13}, R135 (2001).

\bibitem{wuniu} B. Wu and Q. Niu, New J. Phys. {\bf 5}, 104 (2003).


\bibitem{oscfreq} Th. Busch and J. R. Anglin, Phys. Rev. Lett. {\bf 84}, 2298 (2000);
D. J. Frantzeskakis {\it et al.}, Phys. Rev. A {\bf 66}, 053608 (2002); 
V. V. Konotop and L. Pitaevskii, Phys. Rev. Lett. {\bf 93}, 240403 (2003);

\bibitem{last_exp} G.-B. Jo {\it et al.}, Phys. Rev. Lett. {\bf 98}, 180401 (2007);
P. Engels and C. Atherton cond-mat/0704.2427.

\bibitem{atkinson} K. Atkinson,
{\it An introduction to numerical analysis}, John Wiley (New York, 1989).
 
\bibitem{book} L.P. Pitaevskii and S. Stringari,
{\it Bose-Einstein Condensation}, Oxford University Press (Oxford, 2003).

\end{thebibliography}
\end{document}